\documentclass[preprintnumbers, pra, showpacs, floatfix,twocolumn, letterpaper, superscriptaddress]{revtex4-2}
\usepackage{physics}
\usepackage{amsfonts}
\usepackage{amsmath}
\usepackage{amssymb,epsf}
\usepackage{graphicx,epsfig}
\usepackage[margin=1in]{geometry}
\usepackage[colorlinks]{hyperref}
\usepackage{dutchcal}
\usepackage{dsfont}
\usepackage{graphicx}
\usepackage{float}
\usepackage{color,soul}
\usepackage{xcolor}
\definecolor{bluehl}{rgb}{0.6,0.8,1}  % light blue
\sethlcolor{bluehl}

\begin{document}
	\title{Resources of the advantage in quantum illumination: Discord and entanglement}
	\author{Mojtaba Asadollahi}
	\email{mojtaba.77.ma@gmail.com}
	\affiliation{Physics Department, College of Sciences, Shiraz University, Shiraz 71454, Iran}
	\author{Mohammad Hossein Zarei}
	\email{mzarei92@shirazu.ac.ir}
	\affiliation{Physics Department, College of Sciences, Shiraz University, Shiraz 71454, Iran}

\begin{abstract}
We investigate how the quantum advantage in quantum illumination is determined by an interplay between entanglement and discord of the probe state.
In particular, we consider a setup in which the probe is a maximally mixed marginal (MMM) state and the environmental state is completely mixed where
the quantum advantage equals the amount of discord consumed for illumination. Through a conditional extremal analysis, we determine the range of accessible advantages for given values of entanglement and discord. At fixed discord, high entanglement restricts the accessible advantage to high values, making high entanglement sufficient for high advantage. Conversely, at fixed entanglement, low discord limits the advantage to low values, indicating that high discord is necessary for high advantage. These conclusions remain unchanged when alternative correlation measures, including relative entropy and Bures entanglement and geometric discord, are used. The consistency of our results across multiple conceptually distinct measures indicates that the observed resource-advantage relation is not an artifact of a specific quantifier, but a robust feature of the protocol within the family of MMM states. We finally find that in the high-noise regime (near-maximally-mixed state), quantum advantage is, to leading order, a linear function of the initial discord highlighting discord as the key resource for resilience to noise in the protocol.
\end{abstract}

\maketitle

\section{Introduction}\label{sec1}

Since quantum correlations are the key factors for distinguishing classical and quantum systems, quantifying such correlations by suitable measures is one of the most important problems in quantum information theory \cite{chitambar2019quantum,fanchini2017lectures,adesso2016measures,anshu2018quantifying,vedral2003classical}. In particular, quantum entanglement \cite{vedral1997quantifying,horodecki2001entanglement,lee2025quantum,liu2015general,rau2018calculation,yurischev2015quantum} and quantum discord \cite{henderson2001classical,ollivier2001quantum} are two important concepts which have been extensively investigated for characterizing quantum correlations \cite{modi2012classical}. For two-qubit states, it has been shown that entanglement and discord are the same for pure states while they are different for mixed states \cite{qasimi2011comparison}. It implies that both entanglement and discord should be considered for characterizing quantum correlations in the case of two-qubit mixed states.

Besides theoretical importance, quantum correlations are also the main resources of advantage in quantum technologies compared with classical counterparts \cite{ degen2017quantum, ekert1998quantum, steane1998quantum,jozsa2003role,pan2001entanglement}. For example, quantum entanglement in a two-qubit quantum state is a necessary resource for quantum communication protocols such as teleportation and quantum key distribution \cite{pirandola2015advances,sherson2006quantum,vaidman1994teleportation,bennett1993teleporting,jennewein2000quantum,bennett1992quantum,tittel2000quantum}. Multi-partite entanglement in a quantum many-body system is also the main resource for quantum computation in the sense that any computation without sufficient enough entanglement can be classically simulated \cite{amico2008entanglement,nielsen2010quantum}. Conversely, there are quantum information protocols that show quantum advantage while they use discordant separable states as a probe \cite{datta2008quantum,madhok2013quantum,pirandola2014quantum,brodutch2013discord,datta2011quantum}. The main trick behind the above surprise is that such protocols use joint measurements in the entangled bases and it has been shown that such measurements are necessary for achieving a quantum advantage \cite{gu2012observing}. 

An important example of quantum information protocols is quantum illumination where two entangled photons are used as a probe for detection of an object \cite{lloyd2008enhanced}. In particular, one of the photons called signal is sent towards an object and another photon remains for the final joint measurement. Detection is done by performing a joint measurement in the Bell basis on the idler photon and the signal photon that is reflected from the object. Quantum illumination gives an advantage even when the entanglement initially used in the probe state is completely destroyed because of noisy environment. It raises an important question of what is the main characteristic of the initial probe state that leads to the quantum advantage \cite{weedbrook2016discord, kim2023entanglement,yung2020one,jo2021quantum}. In particular, in \cite{weedbrook2016discord}, the authors prove that when the environment is a completely mixed state, the quantum advantage exactly coincides with the amount of discord, called discord of encoding $\delta_{enc}$, that is consumed to encode the presence of the object in the quantum correlation between signal and idler. However, it is shown that when the environment is a thermal state \cite{yung2020one}, equality of discord of encoding and quantum advantage does not always hold except at infinite temperature. Moreover, recent studies on quantum illumination with optical systems show that there are still important ambiguities about the role of discord and entanglement of the probe state which should be carefully considered \cite{kim2023entanglement,yung2020one,jo2021quantum,bradshaw2017overarching}. 

Despite the above studies on the role of entanglement and discord in illumination protocol, a precise and operational delineation of their respective roles remains an open problem. In particular, it is an important task to determine when each resource acts as a necessary or sufficient condition for achieving high quantum advantage. In this work, we address this by performing a conditional extremal analysis across a benchmark family of two-qubit mixed states as the initial state of the probe for quantum illumination while the environment is a completely mixed state. In particular, we consider maximally mixed marginal (MMM) states \cite{maziero2009classical,pozzobom2019preparing,hou2018geometric} which include a wide spectrum of quantum states from Bell states to separable discordant quantum states. It is expected that MMM states (which also called Bell diagonal states) involve important quantum correlations inherent in general two-qubit states \cite{lang2010quantum} because all two-qubit states can be made Bell-diagonal via LOCC. Therefore, they are suitable candidates for studying the role of quantum correlations in the illumination protocol.

First, we confirm that quantum advantage for any given probe state is exactly equal to discord of encoding, consistent with the general theorem proved in \cite{weedbrook2016discord}. Then, we plot diagrams for quantum advantage in terms of entanglement of formation and the standard measure of quantum discord of probe states. We specifically show that there is a broadened relation between quantum advantage and initial entanglement, and between quantum advantage and discord in the sense that there are states with the same initial entanglement or discord but different advantages. It indicates that usable portion of discord for encoding depends on additional structural features of initial probe state. 

To precisely delineate the roles of entanglement and discord, we adopt a conditional extremal analysis. We consider clusters of states with identical advantage and initial discord (or entanglement). Then we find the maximum and minimum initial entanglement (or discord) within each cluster. We show that for a fixed initial discord, the maximum entanglement increases monotonically with the advantage while such an increment is absent for the minimum entanglement. By contrast, for a fixed initial entanglement, we show that the minimum discord (within each cluster) increases monotonically with the advantage while maximum discord shows such an increment only for the low-advantage regime. We finally conclude that high discord is a necessary (and not always sufficient) resource for high advantage while high entanglement is a sufficient (and not necessary) resource for high advantage. 

On the other hand, we notice that there are also other measures for entanglement and discord which can be used in our analysis. Since it has been shown that ordering quantum states in terms of their quantum correlation can be different for different measures \cite{virmani2000ordering}, it is also an important task to repeat our analysis for some well-known and conceptually distinct measures. To this end, we consider relative entropy of entanglement \cite{miranowicz2004comparative,vedral1997quantifying} and Bures measure of entanglement \cite{bromley2014unifying} for MMM states and study geometric discord \cite{dakic2010necessary} as another measure of quantum discord. We show that the above measures behave similarly to entanglement of formation and the standard quantum discord, respectively. This result suggests our conclusion about the role of entanglement and discord as necessary and sufficient resources for quantum illumination with MMM states is not an artifact of a specific measure. Finally, we find a linear dependence, to the leading order, between advantage and initial discord in the high-noise regime where the probe device is extremely noisy, revealing the role of discord as source of resilience of quantum illumination to noise.

The structure of this paper is as follows. In Sec. \ref{sec2}, we give a brief review on quantum illumination with emphasis on the role of joint measurement in quantum illumination protocol. In Sec. \ref{sec3}, we define MMM states and compute quantum entanglement and quantum discord for such states. We use suitable diagrams to visualize the relation between entanglement and discord for MMM states. In Sec. \ref{sec4}, we compute quantum advantage for MMM states and use heat map diagrams to infer the main message of the paper by performing the conditional extremal analysis. Finally, in Sec. \ref{sec 5}, we analytically derive an asymptotically linear dependence between advantage and initial discord in a high-noise regime. 

\section{Quantum Illumination\label{sec2}}
The idea of quantum illumination was proposed by S. Lloyd in his seminal paper \cite{lloyd2008enhanced}. In this section, we use a simple example to explain the main point of the above idea. In particular, while quantum illumination uses two entangled photons as the probe, we use a simple two-qubit state for describing the state of photons. The initial state of the probe is an entangled Bell state in the following form:
\begin{equation}
	|\phi^+\rangle =\frac{1}{\sqrt{2}}(|00\rangle +|11\rangle ),
	\label{eq:bell}
\end{equation}
where $|0\rangle$ and $|1\rangle$ refer to eigenvectors of Pauli operator $\sigma_z$. One of the photons which is called signal is sent towards the target and another one which is called idler remains for the final measurement. We consider a completely mixed state for the environment and therefore the evolution of the state would be described by a depolarizing channel applied on the state and the final state of the signal and idler photons will be in the following form:
\begin{equation}
	\rho^{(0)}_{AB} = \eta \ket{\phi^+}\bra{\phi^+} + (1-\eta)\frac{\mathds{1}_A\otimes \mathds{1}_B}{4},
	\label{eq:werner}
\end{equation}
where $\mathds{1}$ refers to Identity operator and $\eta$ refers to reflectivity factor which is a parameter that describes how much of the wave is reflected by the target. If the target is present, then we would have $\rho^{(0)}_{AB}$ as the total state after reflection. If the target is absent, it means that the signal is lost and we just receive a noise qubit $\rho_E = \mathds{1}_E/2$ in the detector and the total state would be $\rho^{(1)}_{AB}= \rho_B \otimes \rho_E$. Here $\rho_B =\mathds{1}_B /2$ represents the idler qubit.

In order to gain information about the presence or absence of the object, one should perform measurement on idler and signal qubits in a suitable basis. The accessible information is calculated by maximization of the mutual information over all measurement bases. Measurements can be divided into two general sets including single qubit (local) measurements and two-qubit (joint) measurements. As proved in \cite{weedbrook2016discord}, if we perform single-qubit measurement on the idler and then on the signal, the information we can gain about the presence or absence of the object, is equal to the maximum information that we can gain from classical illumination. However, if we perform a two-qubit measurement on both signal and idler in the Bell basis, we will find an advantage which is greater than the classical case.

\begin{figure}
	{\includegraphics[width=0.5\textwidth]{ 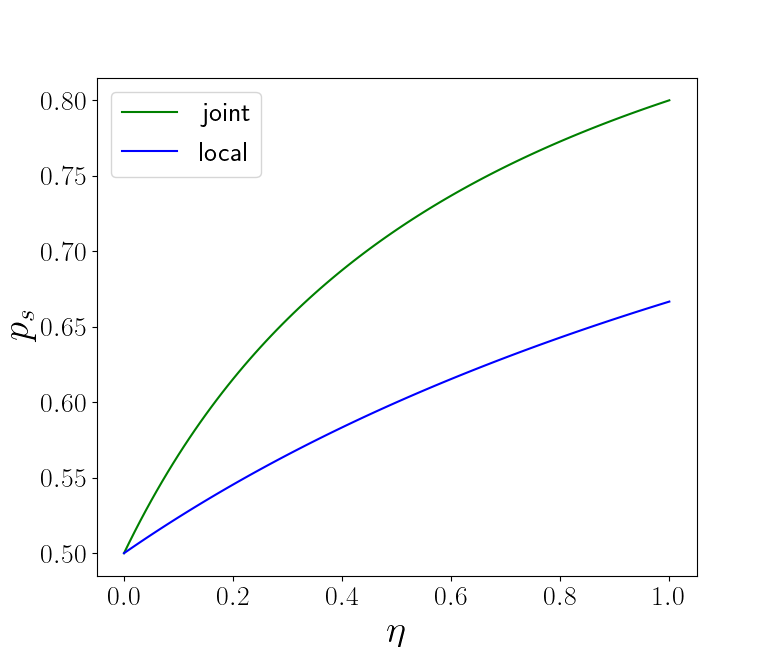}}\centering
	\caption{Probability of a correct guess, $p(there|\text{yes})$, in terms of the reflectivity factor $\eta$. This probability for joint measurement is higher than that for local measurement.}
	\label{frac}
\end{figure}

To compare the difference between the accessible information in the local measurement case and joint measurement case, let us consider two specific bases for measurements including measurement in the Bell basis for joint measurement and in $\sigma_z$ basis for the local measurement case. Notice that if the object is not there, then the state at the detector will be $\rho_B\otimes\rho_E = \mathds{1}\otimes\mathds{1}/4$, which can be written in the following form:

\begin{equation}
	\begin{aligned}
		\frac{\mathds{1}\otimes\mathds{1}}{4} = \frac{1}{4}\bigg[ \ket{\phi^+}\bra{\phi^+} +& \ket{\phi^-}\bra{\phi^-} \\
		+ \ket{\psi^+}\bra{\psi^+}
		& + \ket{\psi^-}\bra{\psi^-}\bigg],
	\end{aligned}
\end{equation}
where $\ket{\phi^\pm} =(\ket{00} \pm \ket{11})/\sqrt{2}$ , $\ket{\psi^\pm} =(\ket{01} \pm \ket{10})/\sqrt{2}$ are orthogonal Bell states. Now, suppose that we perform a joint measurement. Regarding the above form of identity operator in terms of Bell states, there is always a chance with the probability of $1/4$ to gain $\ket{\phi^+}$ and therefore we would say "yes", the object is there. It means that $p_{joint}(\text{yes}|\text{not there}) =1/4$. But if we perform local measurement on the idler first and then on the signal in the $\sigma_z$ basis, we cannot distinguish between $|\phi^+\rangle$ and $|\phi^-\rangle$ and therefore, we would have $p_{local}(\text{yes}|\text{not there}) =1/2$.

Conversely, if the object is there, the final state will be $\rho^{(0)}_{AB}$ and by performing a joint measurement on this state, i.e. Eq. (\ref{eq:werner}), the probability of correctly guessing the presence of the object is in the following form:
\begin{equation}
	p_{joint}(\text{yes}|\text{there}) =  \frac{1+3\eta}{4},
\end{equation}
while if we perform local measurements in the $\sigma_z$ basis, we would have:
\begin{equation}
	p_{local}(\text{yes}|\text{there})  = \frac{1+\eta}{2} .
\end{equation}

Now let us use a Bayesian analysis to calculate $p(\text{there}|\text{yes})=p(\text{yes}|\text{there})p(\text{there})/p(\text{yes})$ to find probability of guessing correctly for both strategies, i.e., joint measurement and local measurement:

$$p_{joint}(\text{there}|\text{yes}) = \frac{1+3\eta}{2+3\eta},$$
\begin{equation}\label{cond2}
	p_{local}(\text{there}|\text{yes})= \frac{1+\eta}{2+\eta}. 
\end{equation}
We plot diagrams related to the above probabilities for different values of $\eta$ for joint and local measurements. As shown in Fig. \ref{frac}, the probability of a correct guess for joint measurement is greater than that of local measurement.\\

\begin{figure}
	{\includegraphics[width=0.47\textwidth]{ 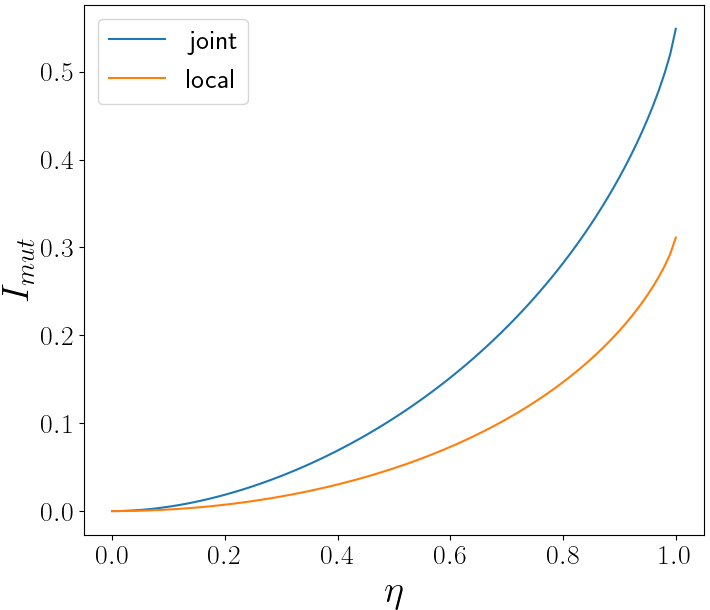}}\centering
	\caption{Mutual information $I_{mut}$ for the case of joint measurement and local measurement in terms of $\eta$. Difference between diagrams for joint measurement and local measurement is called quantum advantage.}
	\label{mut adv}
\end{figure}

In order to quantify the quantum advantage, we should calculate mutual information $I$ between the probability of presence of the object, $X\in\{\text{there},\text{not there}\}$, and the answer in the detector, $K\in\{\text{yes},\text{no}\}$ in the following form:
\begin{equation}
	I_{mut}(X,K) = H(K) - H(K|X), 
	\label{mut}
\end{equation}
where $H(K) = -\sum_{k} p(k)\log p(k)$ is the Shannon entropy for probability distribution related to variable $K$ and 	$H(K|X) =\sum_{x}p(x)H(K|x)$ is the Shannon conditional entropy. Using conditional probabilities derived in Eq. (\ref{cond2}), it is a simple task to compute $H(K)$ and $H(K|X)$.
As shown in Fig. \ref{mut adv}, mutual information for joint measurement is greater than that for local measurement, and the difference between them quantifies quantum advantage in illumination.

We should note that in the above argument we considered specific bases for performing measurements because our goal was only to clarify the concept of quantum advantage. Nevertheless, in general, we should choose arbitrary bases and then calculate accessible information by maximizing the mutual information. In particular, if we replace the initial Bell state used for illumination with a general two-qubit state, optimization for finding the accessible information would not be a simple task. In Sec. \ref{sec4}, we will come back to this point and explain how we can calculate accessible information in a more general case.

\section{Maximally Mixed Marginal states\label{sec3}}
In this paper, we replace the initial Bell state of the probe in the Weedbrook's illumination setup with MMM states as a benchmark family of quantum states while we fix the environmental state to be a completely mixed state.  Before the calculation of the quantum advantage for this more general case, here we give a brief introduction to these states. Specifically, we consider quantum correlations in MMM states by calculating quantum entanglement and quantum discord.

MMM states are defined in the following form \cite{maziero2009classical}:
\begin{equation}
	\rho = \frac{1}{4} \left( \mathds{1}\otimes\mathds{1} + \sum_{i=1}^{3} c_i \sigma_i^{A} \otimes \sigma_i^{B} \right),
	\label{eq:MMM}
\end{equation}
where $|c_i|\le 1$ and $\sigma_1$, $\sigma_2$ and $\sigma_3$ refer to three Pauli operators $\sigma_x$, $\sigma_y$ and $\sigma_z$ respectively. These states are called maximally mixed marginal states because if we trace one subsystem out, the state of the other subsystem will be a completely mixed state. We can also write these states in the matrix form:
\begin{equation}
	\rho =\frac{1}{4}
	\begin{pmatrix}
		1 + c_3 & 0 & 0 & c_1 - c_2\\
		0 & 1 - c_3 & c_1 + c_2 & 0\\
		0 & c_1 + c_2 & 1 - c_3 & 0\\
		c_1 - c_2 & 0 & 0 &1 + c_3
	\end{pmatrix}.  \label{eq:MMMmatrix}
\end{equation}

MMM states include a broad spectrum of quantum states. For example, while for $c_i =0$ the corresponding state is completely mixed state, states with $|c_i|=1 $ correspond to Bell states. In particular, a vector in the form of $(c_1 ,c_2 ,c_3 )$ may be called correlation vector \cite{quan2016steering} in a sense that it represents the most general quantum correlations in two-qubit states. In the following, we introduce some specific classes of quantum states belonging to MMM states. First class is defined by setting $|c_i| =\omega$ and the corresponding states are called Werner states. Second class is defined by setting $c_1 = \alpha$, $c_2 = -\alpha$ and $c_3 = 2\alpha - 1$ and the corresponding states are called $\alpha$-states. Third class is defined by setting $c_1 = 1$, $c_2 = 1-2\beta$ and $c_3 = 2\beta - 1$ and the corresponding states are called $\beta$-states. We will show that these states have special situations among MMM states regarding quantum correlations.

\begin{figure*}
	{\includegraphics[width=0.32\textwidth]{ 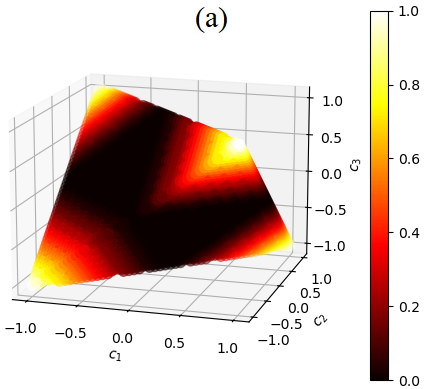}}
	{\includegraphics[width=0.32\textwidth]{ 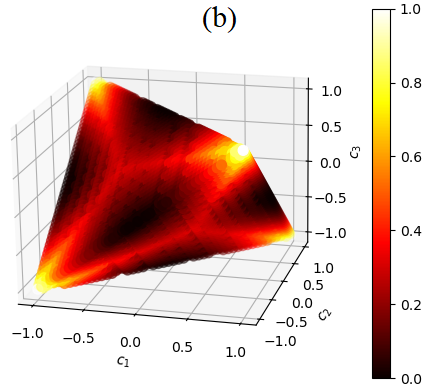}}
	{\includegraphics[width=0.32\textwidth]{ 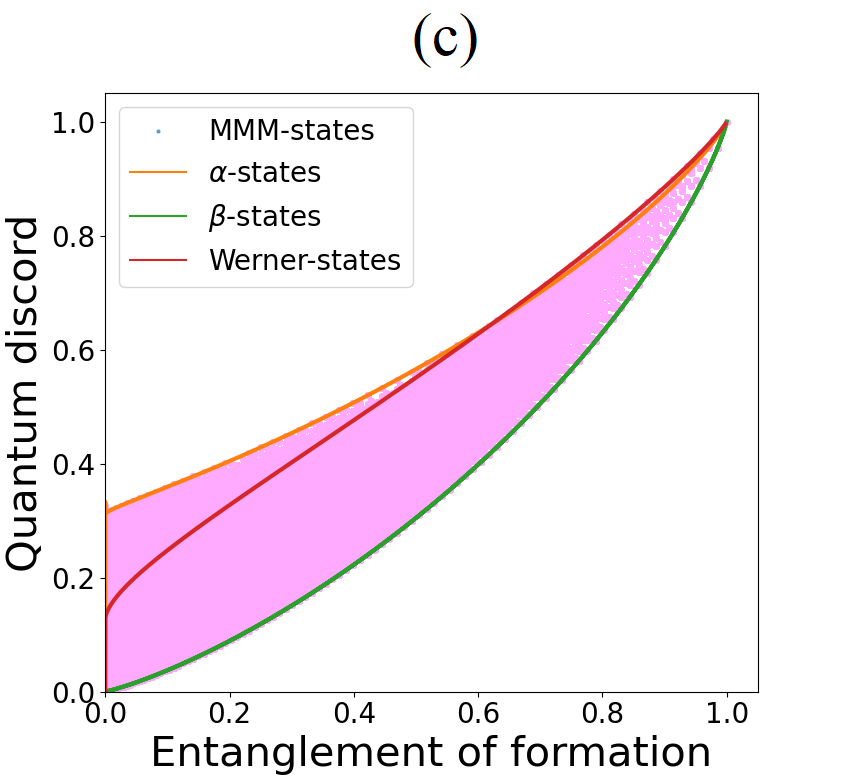}}\centering	
	\caption{a) Entanglement of formation for different values of $c_1$, $c_2$ and $c_3$, is shown by different colors. b) Quantum discord is also shown by different colors. c) Entanglement of formation vs. quantum discord for MMM states (purple dots) is shown. $\alpha$-states (orange line), $\beta$-states (green line) and Werner states (red line) are also plotted. Depending on the entanglement value, in some regions $\alpha$-states and in some regions Werner states are the upper bound. Also, $\beta$-states are the lower bound for the diagram.}\label{entdis}
\end{figure*}

Since we are going to study the role of quantum correlations in the advantage of quantum illumination, here we consider two important measures of quantum correlation including entanglement and quantum discord.  We start by calculating entanglement of formation (EoF) which is related to the concurrence in the following form \cite{wootters1998entanglement,xu2015quantum}:
		\begin{equation}
			\begin{aligned}
				&\text{EoF}(\rho) =\\
				 -&\bigg(\frac{1 + \sqrt{1 - C^2(\rho)}}{2}\bigg)
				\log_2\bigg(\frac{1 + \sqrt{1 - C^2(\rho)}}{2}\bigg)\\
				&- \bigg(\frac{1 - \sqrt{1 - C^2(\rho)}}{2}\bigg)\log_2\bigg(\frac{1 - \sqrt{1 - C^2(\rho)}}{2}\bigg), 
			\end{aligned}
		\end{equation}
where
\begin{equation}
	C(\rho) = \max \bigg\{ 0 , \sqrt{\lambda_1} - \sqrt{\lambda_2} - \sqrt{\lambda_3} - \sqrt{\lambda_4} \bigg\}, 
\end{equation}
and $ \lambda_i$s are the eigenvalues of $\rho \tilde{\rho}$ in decreasing order. $\tilde{\rho}$ is the time-reversed density operator
\begin{equation}
	\tilde{\rho} = (\sigma_y \otimes \sigma_y) \rho^\star (\sigma_y \otimes \sigma_y).
\end{equation}
By using the above formulas, we calculate entanglement of formation for a given $(c_1, c_2 , c_3)$. As shown in Fig. \ref{entdis}(a), in order to illustrate results of calculation, we plot a three-dimensional space with $c_1$, $c_2$ and $c_3$ as axes and we illustrate different values of entanglement by different colors.

Next, we consider quantum discord and compute this quantity for MMM states. We use the standard, mutual-information based method in \cite{ollivier2001quantum} to calculate discord. General definition of quantum discord in a bipartite system needs an optimization over the set of Positive Operator-Valued Measurements (POVMs) on one subsystem. In particular, for MMM states, optimization leads to an analytical formula for discord, as given in \cite{maziero2009classical}:

\begin{equation}
	\delta(\rho_{AB}) = 2 + \sum_{k=1}^{4}\lambda_k\log_2\lambda_k - C(\rho_{AB}),\label{anadis}
\end{equation}
where
\begin{equation}
	C(\rho_{AB}) = \sum_{k=1}^{2}\frac{1 + (-1)^k\zeta}{2} \log_2(1 + (-1)^k\zeta),\label{classical}
\end{equation}
and
$\zeta = \max(|c_1| , |c_2| , |c_3|)$. In the above formula, $\lambda_k$s are eigenvalues of density matrix of MMM states which are in the following form:
\begin{equation}
	\begin{aligned}
		\lambda_1 &= \frac{1}{4}[1-c_1-c_2-c_3], \;\;\lambda_2 = \frac{1}{4}[1-c_1+c_2+c_3], \\
		\lambda_3 &= \frac{1}{4}[1+c_1-c_2+c_3], \;\;\lambda_4 = \frac{1}{4}[1+c_1+c_2-c_3]. 
		\label{eigs}
	\end{aligned}
\end{equation}
So, we can calculate quantum discord for different $c_i$s. As shown in Fig. \ref{entdis}(b), similar to the entanglement, we show the pattern of quantum discord for MMM states on a 3D space. It is important to compare this diagram with Fig. \ref{entdis}(a) for quantum entanglement. In particular, for some values of $c_i$, while there is no entanglement in the corresponding MMM state, there is quantum discord and we call such states discordant separable states. It is shown that MMM states with $|c_i|=1/3$ are maximally discordant separable states. \cite{dakic2010necessary}. 

Comparing entanglement and discord in MMM states can be better shown in a two-dimensional diagram. For this purpose, consider a 2D space where the value of entanglement is shown on horizontal axis and the value of quantum discord is shown on vertical axis. Then for a given $(c_1 , c_2 , c_3 )$ we illustrate the value of entanglement and discord by a point in the 2D diagram. As shown in Fig. \ref{entdis}(c), we plot such diagram for MMM states. It clarifies that discord is not equal to entanglement and there is a broadened relation between discord and entanglement in MMM states. It means that there are MMM states with the same entanglement (discord) which have different discord (entanglement). We have specifically plotted curves corresponding to $\alpha$-states, $\beta$-states and Werner states. As shown in Fig. \ref{entdis}(c), $\beta$-states are the lower bound of the diagram. It means that for MMM states with a given entanglement (discord), $\beta$-states have minimum discord (maximum entanglement). On the other hand, the upper bound of the diagram is different for states with low correlation and high correlation. In the  low correlation regime, $\alpha$-states are the upper bound in a sense that for MMM states with a given entanglement (discord), $\alpha$-states have maximum discord (minimum entanglement) and in the high correlation regime, Werner states are the upper bound. 

The broadened relation between entanglement and discord in two-qubit mixed states shows that quantum correlations cannot be uniquely captured by one of these quantifiers and both are important. 

It means that quantum correlation has different aspects which should be quantified by different measures. Also, quantum correlations are the main resource for quantum advantages in quantum information protocols. On this basis, it is important to consider how these two quantifiers play roles in providing quantum advantage in a quantum information protocol. In the next section, we consider this problem for quantum advantage in illumination protocol when we use MMM states as initial probe.

\section{Role of entanglement and discord in quantum advantage for MMM states\label{sec4}}
After reviewing the quantum illumination protocol in addition to quantum correlations in MMM states, now we are ready to consider our main problem by calculating the quantum advantage in the illumination protocol when we use an MMM state as the probe state and a completely mixed state as the environmental state. Our main goal of considering such setup is to study the role of entanglement and discord of the probe state in the quantum advantage. In particular, notice that as shown in the previous section, when the probe is in a Bell state both entanglement and discord are maximal. However when probe is in a mixed state, determining the contribution of entanglement and discord would be challenging because there is a broadened relation between entanglement and discord as shown in Fig. \ref{entdis}(c). Accordingly, it is the main goal of our paper to clarify the contribution of the above measures in the quantum advantage. Considering a two-qubit mixed state as a probe state is practically reasonable because the probe state may be generated by a noisy device and therefore, it can be mixed due to noise. Notice that this noise is different from the environmental noise, where we have fixed the environmental state to a completely mixed state for arbitrary probe states.

In order to start our study, we recall that, in Sec. \ref{sec2}, we introduced the general idea of the calculation of quantum advantage. In order to quantify the quantum advantage, we should calculate the accessible information \cite{holevo2002capacity}. This quantity has to be computed between the two states $\rho^{(0)}$ and $\rho^{(1)}$ corresponding to presence and absence of the object \cite{weedbrook2016discord,bradshaw2017overarching}. It is equal to maximum value of mutual information \cite{bradshaw2017overarching}:

\begin{equation}
	A (\rho^{(0)},\rho^{(1)})= \max_M I_{\text{mut}}(X , K_M),
\end{equation}
where $I_{\text{mut}}$ is the mutual information, $M$ is the set of positive operator-valued measures (POVMs), $K_M$ is a random variable related to measurement outcome and $X$ is a random variable which is related to presence or absence of the object.

On the other hand, it has been shown that accessible information is bounded by the Holevo information \cite{fuchs1994ensemble,fuchs1996distinguishability}:
\begin{equation}
	\begin{split}
		A (\rho^{(0)} , \rho^{(1)}) \le \chi (\rho^{(0)} , \rho^{(1)}),
		\label{eq:holevo}
	\end{split}
\end{equation}
where $\chi$ refers to Holevo information which is defined in the following form:
\begin{equation}
	S \left( \sum_{x=0}^{1} p_x \rho^{(x)} \right)  - \sum_{x=0}^{1} p_x S(\rho^{(x)}),
\end{equation}
where $p_x$ is the prior probability about the presence or absence of the object, and $S(.)$ denotes the von Neumann entropy. The inequality turns to equality if and only if $\rho^{(0)}$ and $\rho^{(1)}$ commute \cite{fuchs1994ensemble}. Fortunately, for MMM states, $\rho^{(0)}$ and $\rho^{(1)}$ commute and therefore we calculate Holevo information to evaluate the amount of quantum advantage. 

For MMM states we write quantum advantage as difference between Holevo information for quantum illumination with joint-measurement and that with local measurement:
\begin{equation}
	QA = \chi_q(\rho^{(0)} , \rho^{(1)}) - \chi_c(\rho^{(0)}_c, \rho^{(1)}_c),
	\label{eq:QA}
\end{equation}
where $\chi_q =S(\bar{\rho} ) - p_0S(\rho^{(0)}) - p_1S(\rho^{(1)})$ refers to Holevo information for quantum illumination with joint measurement and $\chi_c =S(\bar{\rho_c}) - p_0 S(\rho^{(0)}_c) - p_1S(\rho^{(1)}_c)$ refers to Holevo information for conventional illumination with local measurement where $\rho^{(0)}_c$ and $\rho^{(1)}_c$ refer to the density matrices of the whole system when a local measurement is performed on the idler first, followed by measurement of the signal where a minimization on different bases of measurements is done. In \cite{weedbrook2016discord} it is shown that in this strategy, the accessible information equals the maximum accessible information that one can gain in conventional illumination (with local measurement).
\begin{figure}
	{\includegraphics[width=0.4\textwidth]{ 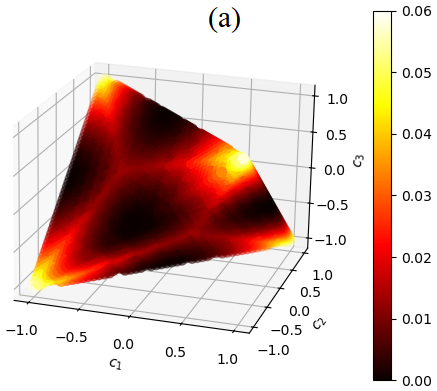}}
	{\includegraphics[width=0.45\textwidth]{ 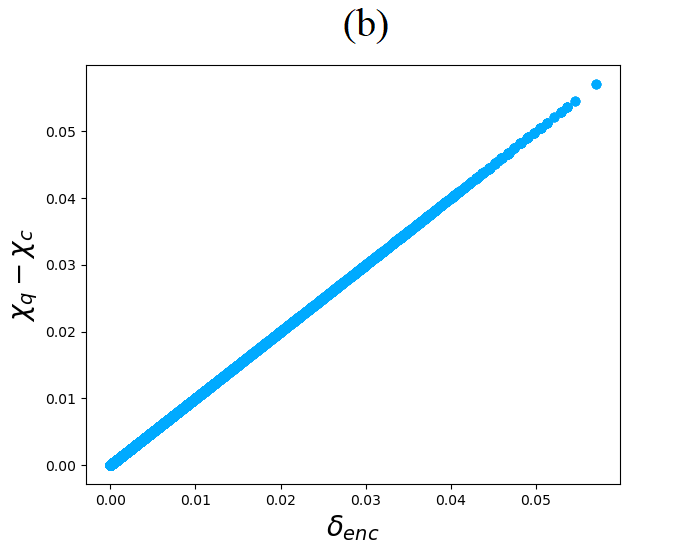}}
	\caption{a) Quantum advantage of MMM-states for different values of $c_1$, $c_2$ and $c_3$ is shown by different colors. b) Quantum advantage vs. discord of encoding. It shows that they are equal for any given MMM state. The reflectivity factor here is $\eta=0.5$.}\label{holevodenc}
\end{figure}

Using the above formula, it is a simple task to compute quantum advantage for different MMM states in terms of  $c_1$, $ c_2$ and  $c_3$ as shown in Fig. \ref{holevodenc}(a). Accordingly, when the state of the environment is completely mixed, the best states for quantum illumination are Bell states as expected \cite{ray2019maximum,yung2020one}. Also, we can compare the pattern of quantum advantage with the patterns of initial entanglement and discord, i.e., Fig. \ref{entdis}(a) and Fig. \ref{entdis}(b), plotted in the previous section. One can simply conclude that the pattern of quantum advantage is well matched with the pattern of initial discord while it is different from the entanglement pattern. In particular, there are initial separable MMM states (without entanglement) which show quantum advantage.

On the other hand, relation between quantum advantage and initial discord has also been investigated in \cite{weedbrook2016discord} where it is proposed that quantum advantage is literally the same as a quantity called discord of encoding which is the amount of discord consumed to encode the presence of the object in the initial quantum correlation. It is defined in the following form:
\begin{equation}
	\delta_{enc} = p_0 \delta^{(0)}(\rho_{AB}) - \delta(\bar{\rho}).
	\label{eq:denc}
\end{equation}
Here, $\delta_{enc}$ is the discord of encoding, $\delta^{(0)}$ is the discord corresponding to $\rho^{(0)}$, and $\delta(\bar{\rho})$ is the discord corresponding to an average state $\bar{\rho} = p_0\rho^{(0)} +  p_1\rho^{(1)}$. Given this, and by using Eq. (\ref{anadis}), we compute discord of encoding for all MMM states and accordingly, we plot quantum advantage in terms of discord of encoding for all $c_i$s. As shown in Fig. \ref{holevodenc}(b), all states fall on a line such that $QA=\delta_{enc}$. 

\begin{figure*}
	{\includegraphics[width=0.315\textwidth]{ 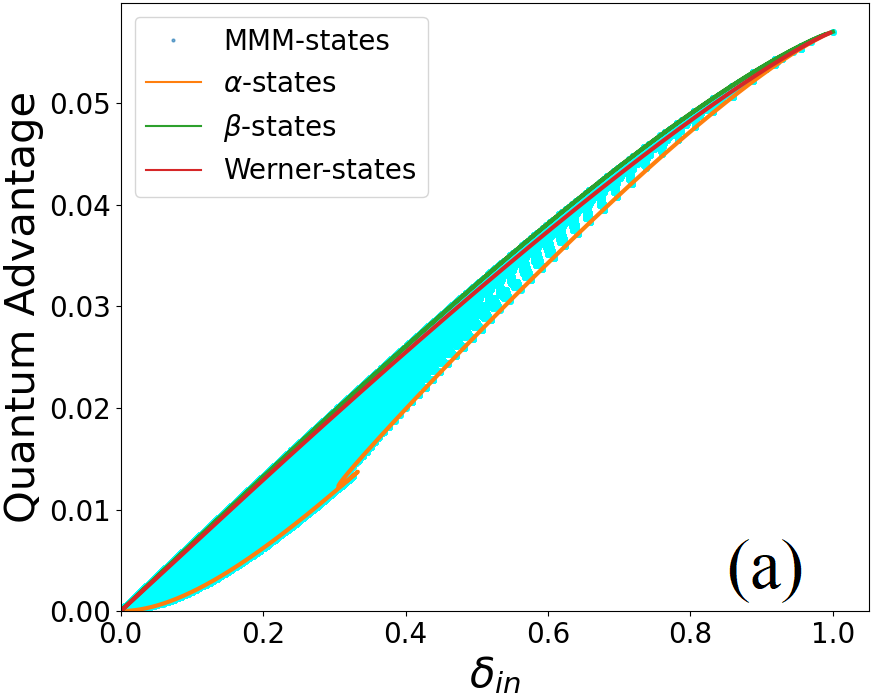}}
	{\includegraphics[width=0.33\textwidth]{ 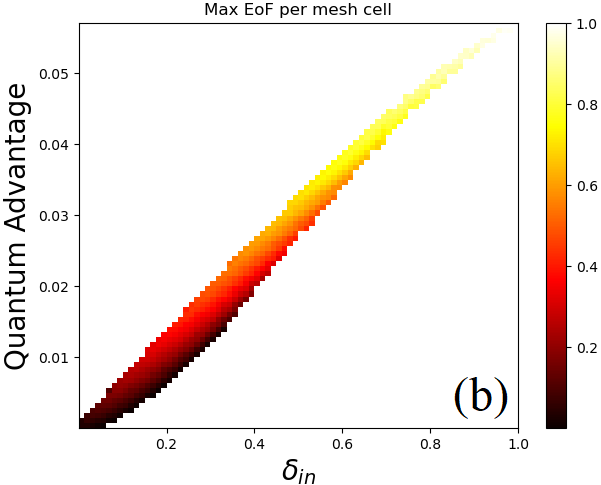}}
	{\includegraphics[width=0.325\textwidth]{ 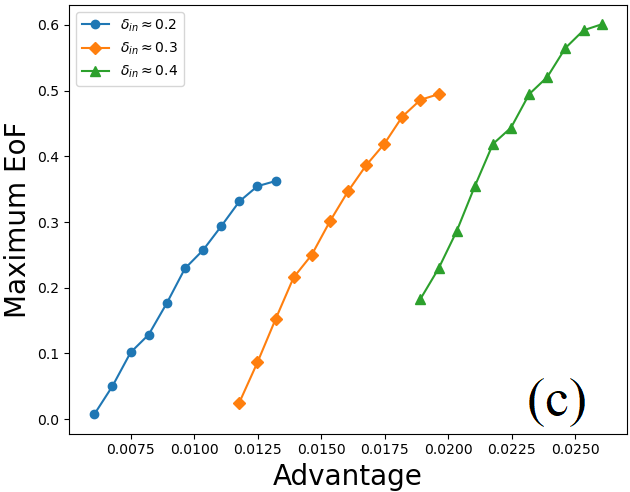}}
	\caption{\textbf{Quantum advantage versus initial discord for $\eta=0.5$.}	
		a) Light blue dots are MMM states. $\alpha$-states, $\beta$-states and Werner states are plotted as orange, green and red line respectively. $\alpha$-states are the lower bound and $\beta$-states are the upper bound. There is a small horn near  $\delta_{in}=0.33$ in the diagram which is related to the separable states. b) A similar diagram for only entangled states where the small horn in (a) disappears. Among a cluster of states in each mesh cell, states with maximum entanglement are chosen. Precision of the mesh is 1/80. The color bar shows the maximum entanglement among the states of each cluster. c) Corresponding to three columns in (b), for three given values of the initial discord, maximum entanglement vs. the quantum advantage is plotted. Entanglement increases with increasing quantum advantage. }\label{disdenc}
\end{figure*}

The above result shows that MMM states with the same quantum advantage have the same value of discord of encoding. It does not mean that all states with the same advantage have necessarily the same initial discord. In order to clarify this point, we plot quantum advantage in terms of initial discord for all MMM states. As shown in Fig. \ref{disdenc}(a), it leads to a particular diagram in a sense that relation between quantum advantage and initial discord is broadened. In other words, there are states with the same initial discord that lead to the different quantum advantage. Since quantum advantage is the same as discord of encoding, our result implies that for states with the same initial discord, there should be another important factor which determines how much discord is consumed for illumination. 

With this aim, first let us plot curves corresponding to $\alpha$-states, $\beta$-states and Werner states in the above diagram. As shown in Fig. \ref{disdenc}(a), these states are bounds of the diagram. In particular, $\beta$-states are the upper bound and $\alpha$-states are the lower bound. Recall from Fig. \ref{entdis}(c) for discord-entanglement relation that for a given initial discord, $\beta$-states were states with maximum entanglement and $\alpha$-states were states with minimum entanglement. It implies, for a given initial discord, there is a correlation between the amount of initial entanglement and the quantum advantage in a sense that $\beta$(and $\alpha$)-states have maximum (and minimum) advantage and maximum (and minimum) initial entanglement. It seems that entanglement should be that factor which determines how much discord is consumed for illumination. However, there is an ambiguity about Werner states. While according to Fig. \ref{entdis}(c), Werner states are low-entangled for a given discord, they are high-advantage according to Fig. \ref{disdenc}(a). It shows that the role of entanglement in the quantum advantage should be considered by a more reliable analysis. 

\begin{figure*}
	{\includegraphics[width=0.31\textwidth]{ 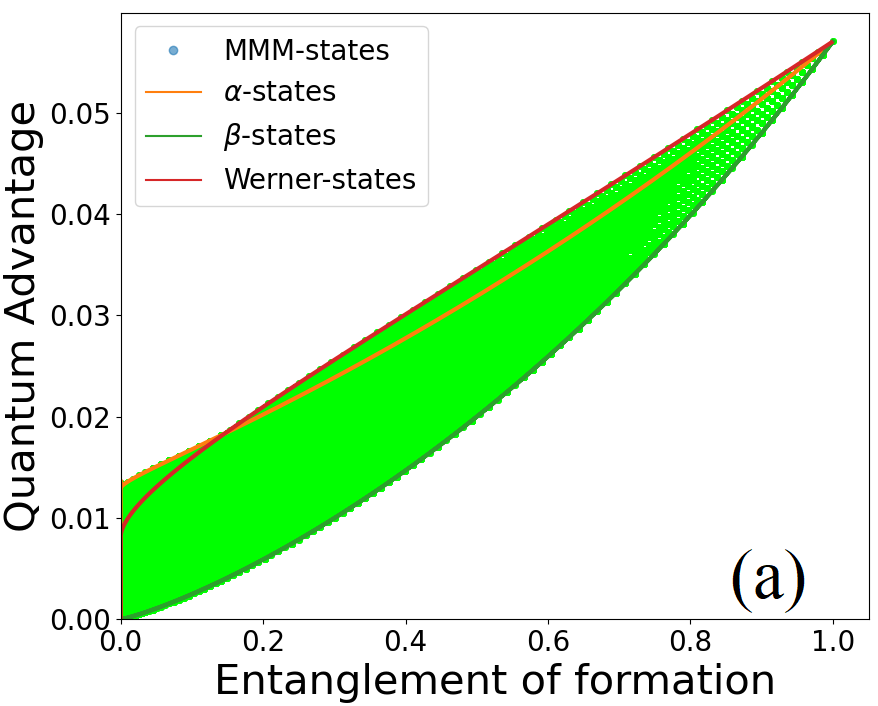}}
	{\includegraphics[width=0.33\textwidth]{ 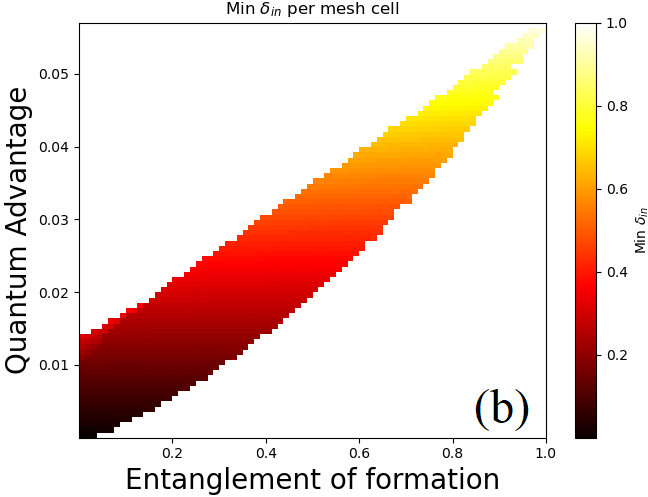}}
	{\includegraphics[width=0.32\textwidth]{ 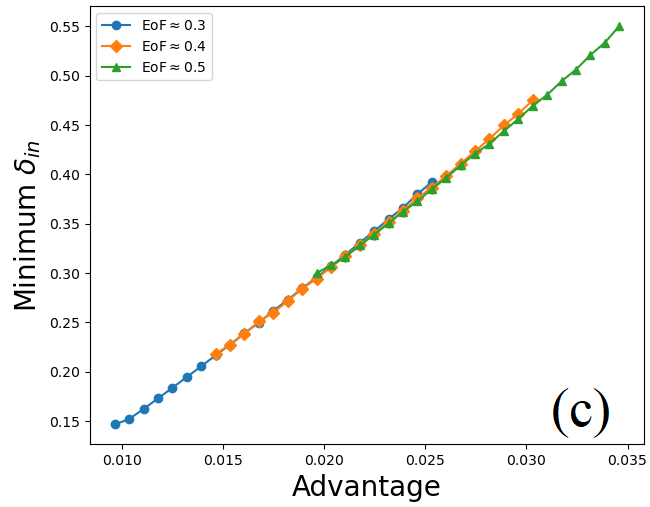}}
	\caption{\textbf{Quantum advantage versus initial entanglement for $\eta=0.5$.} a) Green dots are MMM states. $\alpha$-states, $\beta$-states and Werner states are plotted as orange, dark green and red lines, respectively. b) Among the states of each mesh cell, the state with minimum discord is chosen. Precision of the mesh is 1/80. Minimum values of initial discord for different mesh cells are denoted by different colors. It scales monotonically with the advantage. c) For three given values of initial entanglement, the minimum discord vs. quantum advantage is plotted. By increment of the quantum advantage, minimum discord monotonically increases. }\label{entdenc}
\end{figure*}
\begin{figure}
	{\includegraphics[width=0.425\textwidth]{ 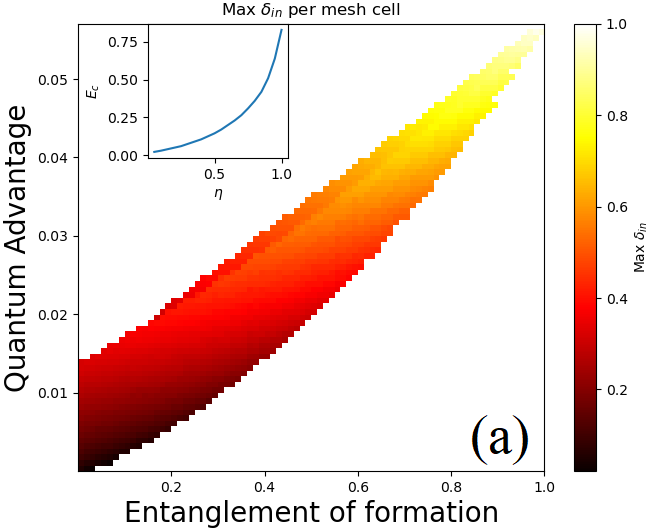}}
	{\includegraphics[width=0.425\textwidth]{ 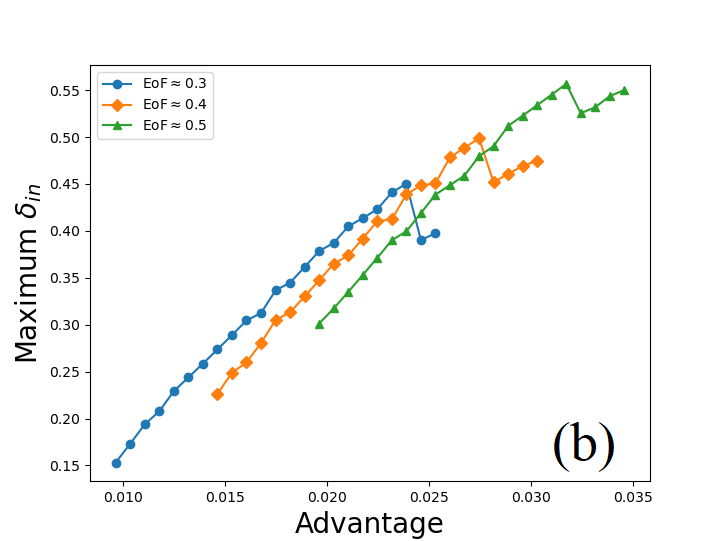}}
	\caption{\textbf{Quantum advantage versus entanglement for maximum discord.} a) Among the states of each mesh cell, the state with maximum discord is chosen. Precision of the mesh is 1/80. The inset shows value of entanglement at the transition point, where the role of Werner state and $\alpha$ states is switched, versus the reflectivity $\eta$. b) For three given values of initial entanglement, the maximum discord (of each mesh cell) vs. quantum advantage is plotted. We can see that it is not an increasing function for all values of the quantum advantage. }\label{entdencmax}
\end{figure}

We perform a conditional extremal analysis for finding relation between advantage and initial entanglement for a given initial discord. First we limit our study to only entangled MMM states. Then, we consider clusters (considered in a mesh cell in Fig. \ref{disdenc}(b)) of entangled MMM states with identical quantum advantage and initial discord. We find a specific state with maximum entanglement (upper boundary) in each cluster. In Fig. \ref{disdenc}(b), we plot a heat map diagram where the values of maximum entanglement for different clusters are illustrated by different colors. As seen in the figure, for a given initial discord, the quantum advantage is an increasing function of value of the maximum entanglement. For more clarification, in Fig. \ref{disdenc}(c), we have shown this behavior for some specific values of initial discord.  According to the above increasing behavior, we are able to determine what range of advantages is accessible for a given value of entanglement. In particular, notice that a state with entanglement $E$ can only belong to bins whose upper boundary is not less than $E$. Hence a state whose entanglement lies above the upper boundaries of the low-advantage bins is necessarily a high-advantage state. In other words, for a fixed discord, while having low entanglement yields access to a wide range of advantages including low and high advantages, for high-entangled states the range of advantage becomes restricted to high values. It means that high entanglement is sufficient for high advantage in a sense that one would never reach low advantage by having sufficiently high entanglement.  

By contrast, we recall that there are separable states which show quantum advantage. Therefore, it is simply concluded that if we consider value of minimum entanglement for each cluster in the above analysis, there would not be an increasing behavior in terms of the advantage. Therefore, high entanglement is not a necessary resource for quantum advantage. It means that for a given discord, there are states with low entanglement which show high advantage. Werner states are clear examples of such states.

We would also like to emphasize that in Fig. \ref{disdenc}(a), a small horn is seen at quantum advantage at $\delta_{in} = 0.33$ while it disappears in Fig. \ref{disdenc}(b) where we have only plotted entangled states. In other words, the above horn is related to separable discordant states and transition from separable states to entangled states has a signature in the quantum advantage diagram.

In the next step, we perform a similar conditional extremal analysis for studying relation between quantum advantage and initial discord for a given entanglement. As shown in Fig. \ref{entdenc}(a), we plot a diagram for the quantum advantage in terms of initial entanglement. It shows a more broadened relation compared to the advantage-discord relation. Furthermore, the role of $\beta$-states and $\alpha$-states has been reversed. $\beta$-states are the lower bound in the advantage-entanglement relation. $\alpha$-states and Werner states are the upper bounds in the low correlation and high correlation regimes, respectively. Recall from Fig. \ref{entdis}(c) where $\beta$-states have minimum discord for a given entanglement while $\alpha$-states and Werner states have maximum discord for a given entanglement. This implies that, for a given entanglement, high advantage also correlates with high discord. 

In order to clarify the above point, we consider clusters of states with identical advantage and entanglement and then compute discord for states belonging to each cluster. First we find a state with minimum discord (lower boundary) in each cluster. As shown in Fig. \ref{entdenc}(b), we plot a heat map diagram where value of minimum discord for different clusters is shown by different colors. As seen in the mentioned figure, for a given entanglement, values of the minimum discord are an increasing function of the advantage. In Fig. \ref{entdenc}(c), we have shown this behavior for some specific values of entanglement. For a fixed entanglement, the minimum discord within each advantage bin increases with the advantage. A state with discord $\delta$ can only belong to bins whose lower boundary does not exceed $\delta$. Hence a state whose discord lies below the lower boundaries of the high-advantage bins cannot reach high advantage. In other words, low discord makes the range of accessible advantages restricted to low values. It simply implies that high discord is necessary for high advantage.

On the other hand, we can also find states with maximum discord in each cluster corresponding to a given advantage and entanglement. In Fig. \ref{entdencmax}(a), we plot another heat map diagram where values of maximum discord for different clusters are illustrated by different colors. Importantly, it is seen that the maximum discord is not always an increasing function of the quantum advantage. This is clearer in Fig. \ref{entdencmax}(b), where we have shown this behavior for some specific values of entanglement. It implies that high discord is not always a sufficient resource for high advantage. In particular, in the high advantage regime near the upper bound of the diagram, there are states with lower discord which show high advantage. However, in the low-advantage regime, increasing behavior is seen and it means that high discord is a necessary and sufficient resource for high advantage in the above regime.

It is also interesting to compare advantage-entanglement diagram in Fig. \ref{entdenc}(a) with discord-entanglement diagram in Fig. \ref{entdis}(c). In particular, according to Fig. \ref{entdis}(c), $\alpha$-states had maximum discord for most values of entanglement, while according to the diagram of Fig. \ref{entdenc}(a), Werner states have maximum advantage for most values of entanglement. This is exactly why advantage does not show an increasing behavior with discord in the high advantage regime. In other words, peaks in diagrams Fig. \ref{entdencmax}(b) are related to Werner states. We should also emphasize that we have plotted the diagram Fig. \ref{entdencmax} for reflectivity factor $\eta=0.5$. In particular, there is a transition point in Fig. \ref{entdenc}(a) where  the role of $\alpha$ states and Werner states as the upper bound is interchanged. This transition point depends on the reflectivity $\eta$ in a sense that by increasing $\eta$, this transition occurs at higher entanglement values. In Fig. \ref{entdencmax}(a), the inset shows the transition point in terms of $\eta$. Accordingly, a domain of the heat map diagram in which high discord is not a sufficient resource for high advantage depends on $\eta$.
\begin{figure}
	{\includegraphics[width=0.235\textwidth]{ 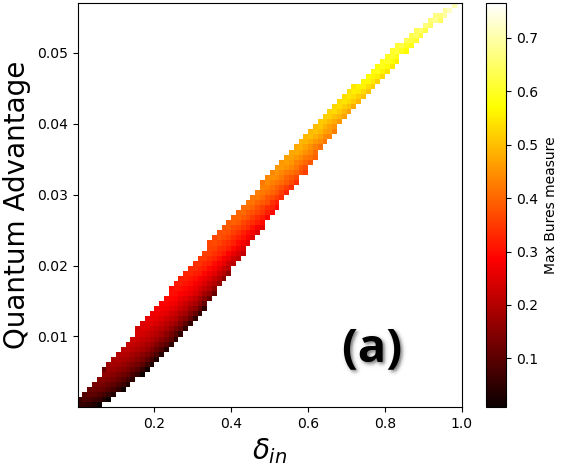}}
	{\includegraphics[width=0.235\textwidth]{ 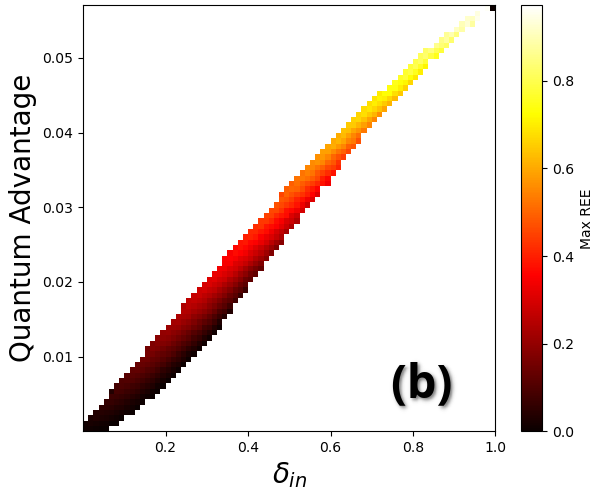}}
	{\includegraphics[width=0.235\textwidth]{ 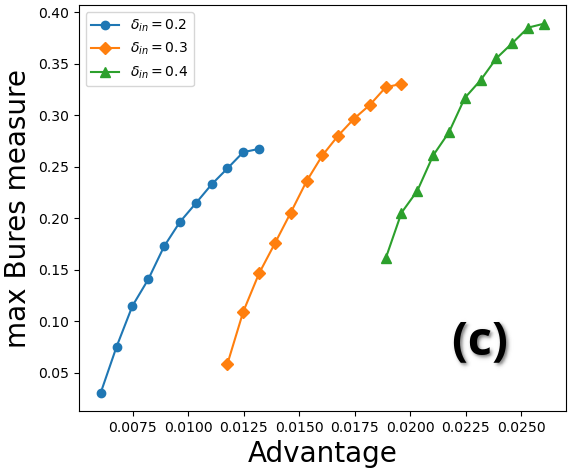}}
	{\includegraphics[width=0.235\textwidth]{ 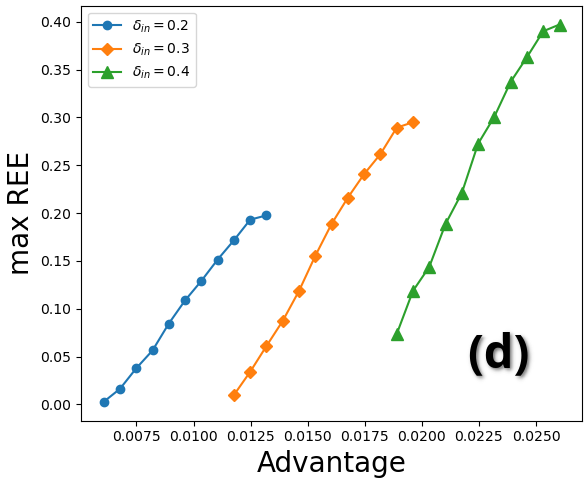}}
	\caption{Same as Fig. \ref{disdenc}(b) and (c), but here for entanglement, in (a) Bures measure of entanglement is calculated, and in (b) relative entropy of entanglement. (c) and (d) are for fixed discords where maximum entanglement vs quantum advantage are plotted for (a) and (b) diagrams respectively. }\label{2entmeas}
\end{figure}
In summary,  using the above arguments on the diagrams, we conclude that while high entanglement is a sufficient (and not necessary) resource for high advantage, quantum discord is a necessary (and not always sufficient) resource for high advantage. Nonetheless, this result is restricted to the  measures we used for entanglement and discord. In order to consider generality of the result, we should consider other measures of quantum correlations. Fortunately, there are simple formulas for different measures of entanglement and discord in MMM states. Accordingly, here we repeat our analysis for relative entropy of entanglement $E_{REE}$ \cite{miranowicz2004comparative,vedral1997quantifying} and Bures measure of entanglement $E_B$ \cite{bromley2014unifying} which have a simple formula for MMM states. According to the formula presented in \cite{quan2016steering}, $E_{REE}$ is written as
\begin{equation}
	E_{REE}= 1+ \lambda_{\text{max}}\log_2\lambda_{\text{max}} + (1-\lambda_{\text{max}})\log_2(1-\lambda_{\text{max}}),
\end{equation}
and $E_B$ for MMM states as
\begin{equation}
	E_B =  \sqrt{2 \bigg( 1-\sqrt{\frac{1}{2}+\sqrt{\lambda_{\text{max}}(1-\lambda_{\text{max}})}} \bigg)},
\end{equation}
where $\lambda_{\text{max}}$ is the largest eigenvalue in Eqs. (\ref{eigs}). 

\begin{figure}
	{\includegraphics[width=0.235\textwidth]{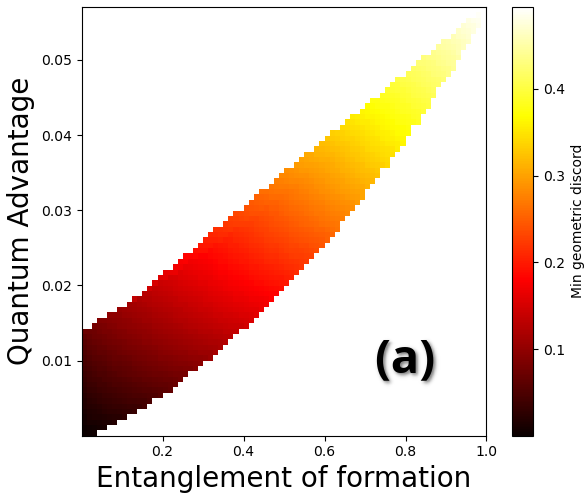}}
	{\includegraphics[width=0.235\textwidth]{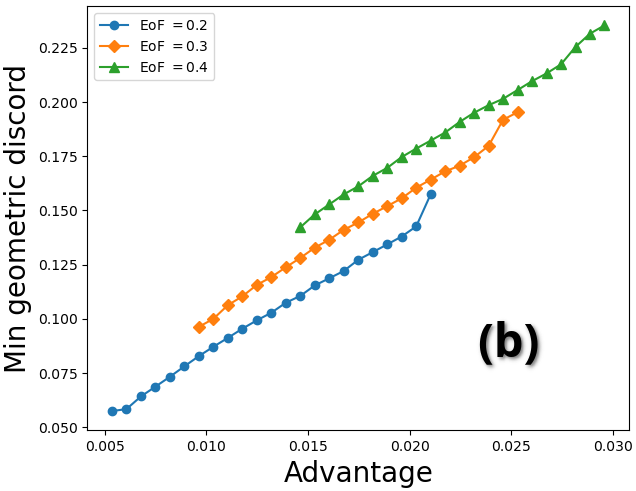}}
	\caption{Same as Fig. \ref{entdenc}(b) and (c), but here a) geometric discord is calculated instead of mutual information based discord. b) For three fixed entanglement, geometric discord vs quantum advantage is plotted. }\label{gdis}
\end{figure}

These measures are conceptually distinct from entanglement of formation. For example, it is shown that relative entropy of entanglement and entanglement of formation lead to different orderings of quantum states in terms of their entanglement \cite{virmani2000ordering}. As shown in Fig. \ref{2entmeas}, for a fixed value of discord, the maximum value of the above measures shows an increasing behavior in terms of quantum advantage. It confirms our previous result that high entanglement is a sufficient resource for high advantage.

We also consider geometric discord \cite{piani2012problem} as another measure of quantum discord in MMM states by the following formula:
\begin{equation}
	GD= \frac{1}{4}\bigg( c_1^2 + c_2^2 + c_3^2 - \max(c_1^2 , c_2^2 , c_3^2 ) \bigg)
\end{equation}
where $c_i$s are parameters of the MMM states in Eq. (\ref{eq:MMM}).

This measure is also conceptually different from standard discord where it can increase under some local operations \cite{piani2012problem}. Interestingly, as shown in Fig. \ref{gdis}, the minimum value of this quantity shows the expected increasing behavior that we saw for standard quantum discord. These results show that multiple distinct measures of entanglement and discord provide the same description of role of entanglement and discord in quantum advantage. The convergence of these distinct measures suggests that our findings reflect a robust feature of the protocol rather than an artifact of a particular measure.

\section{Resilience to noise in the high-noise regime\label{sec 5}}
As shown in the previous section, in the low-advantage regime, high discord is a necessary and sufficient resource for high advantage. In particular, if we consider MMM states which are very close to the completely mixed state, we expect that quantum discord can capture essential features related to quantum advantage. On the other hand, MMM states near the completely mixed state can be regarded as quantum states which are generated by a highly noisy device. In this regard, quantum illumination with such states can be regarded as quantum illumination in a high-noise regime. Such a study can in particular reveal the role of discord as the resource for resilience of quantum advantage to noise.

In the high noise regime (e.g. using a highly noisy device or infinitesimal reflectivity of the target), the initial state is very close to a completely mixed state and therefore the  correlation vector approaches zero. It is clear that both quantum advantage and initial discord go to zero when correlation vector approaches zero. However, the ratio of quantum advantage to initial discord might go to zero, a finite value, or infinity which show sub-linear, linear and super-linear relation of quantum advantage in terms of initial discord, respectively. A linear or sub-linear relation can be regarded as evidence for this point that quantum discord is the source of resilience to noise in quantum illumination. 
So, we give an analytical argument to compute the following limit: 
\begin{equation}
	\lim\limits_{(c_1,c_2,c_3)\rightarrow 0} \frac{QA}{\delta_{in}}
	\label{eq:lim}
\end{equation}
Here, $\delta_{in}$ and $QA$ refer to initial discord and quantum advantage, respectively. For calculating the discord of encoding $\delta_{enc}= p_0 \delta^{(0)}-\bar{\delta}$, we should consider discord for $\rho^{(0)}$ and $\bar{\rho}$ which are the following form:
\begin{equation}
	\begin{aligned}
		\rho^{(0)} &=\frac{1}{4}\left( \mathds{1}\otimes\mathds{1} + \eta\sum_{i=1}^{3}c_i\sigma_i^A\otimes\sigma_i^B \right)\\
		\bar{\rho} &= \frac{1}{4}\left( \mathds{1}\otimes\mathds{1} + p_0\eta\sum_{i=1}^{3}c_i\sigma_i^A\otimes\sigma_i^B \right).\label{cal43}
	\end{aligned}
\end{equation}
Then we use Eq. (\ref{anadis}) to find an expression for quantum discords corresponding to the above states. As proved in detail in Appendix \ref{appen}, if we use a Taylor expansion for the logarithm terms, the quantum discords take the following form to leading order as $(c_1,c_2,c_3)\rightarrow 0$
		
		\begin{equation}
			\delta_{in}
			=
			\frac{1}{2\ln 2}(c_2^2+c_3^2)
			+o(c_2^2+c_3^2),
			\label{a31}
		\end{equation}
		and 
		\begin{equation}
			p_0\delta^{(0)}-\bar{\delta}
			=
			\frac{p_0\eta^2(1-p_0)}{2\ln 2}
			(c_2^2+c_3^2)
			+o(c_2^2+c_3^2).
			\label{a30}
		\end{equation}
		
		where $o(c_2^2+c_3^2)$ denotes that the remaining terms vanish faster than $c_2^2+c_3^2$ itself. Combining these, the limit in Eq. (\ref{eq:lim}) is of the form
		\begin{equation}
			\lim\limits_{(c_1,c_2,c_3)\rightarrow 0}\frac{p_0\delta^{(0)} - \bar{\delta}}{\delta_{in}} = p_0 \eta^2 (1-p_0), \label{limresult}
		\end{equation}
		
		This shows that, asymptotically as the correlation vector approaches zero (i.e. in the high-noise regime), quantum advantage is related to initial discord by
		\begin{equation}
			\delta_{enc} = p_0 \eta^2 (1-p_0) \delta_{in} + o(\delta_{in}) \qquad ||c||\rightarrow 0. \label{limitt}
		\end{equation}
		Here, also, $o(\delta_{in})$ denotes a remainder that vanishes faster than $\delta_{in}$ itself. Equation (\ref{limitt}) is therefore a local, leading-order relation, valid when the correlation coefficients $c_i$
		are small compared to unity (equivalently, when the initial state is close to the maximally mixed state); it should not be read as an exact identity at finite  $c_i$. Away from this regime, higher-order corrections computed in Appendix \ref{appen} (the cubic and quartic terms in Eqs. (\ref{a23})–(\ref{a25})) contribute and the linear relation no longer holds exactly. Within its regime of validity, this local linear behavior still supports the interpretation of quantum discord as a source of resilience of quantum illumination to noise in the high-noise limit.
%This linear behavior reveals the role of quantum discord as the source of resilience of quantum illumination to noise. 

\section{Conclusion}
In spite of the simple structure of two-qubit states, quantifying quantum correlations in general two-qubit mixed states needs more clarification. In particular, using such states as resource for different quantum information protocols sheds light on this problem. Here, we used MMM states as the probe for quantum illumination protocol while the environment is in a completely mixed state. Then, we computed entanglement of formation and standard quantum discord of the initial MMM state and showed that the above measures of quantum correlation do not have a straightforward relation to the final advantage. We found a broadened relation between quantum advantage with entanglement and discord and introduced suitable diagrams for illustrating the above broadened relations. We identified $\alpha$-states, Werner states and $\beta$-states as bounds for these diagrams with maximum or minimum advantages. We then used heat map diagrams to visualize relation between quantum advantage with both entanglement and discord. Using a conditional extremal analysis, we showed that high discord is a necessary (but not always sufficient) resource for high advantage while high entanglement is a sufficient (but not necessary) resource.  Also, in the high-noise regime of the probe device where the probe state is near a completely mixed state, we observed that discord plays an essential role where high discord is both a necessary condition and a sufficient condition. In this regime, we found an asymptotically linear dependence, to the leading order, of the advantage on the initial discord. This linear relation holds locally, in the limit of vanishing correlation vector (high-noise regime). It indicates that discord acts as a robust quantum resource that maintains a linear contribution to the performance of the illumination protocol even when entanglement and other quantum features are heavily suppressed by noise.

We also checked validity of our results by considering other measures of entanglement and discord. In particular, we showed that in spite of the conceptual distinction between measures that we consider, they exhibit a qualitatively consistent monotonic relationship with the quantum advantage within the family of MMM states. The convergence of these distinct measures on the same qualitative trend suggests that our findings reflect a robust feature of the protocol rather than an artifact of a particular quantifier. We emphasize that although MMM states involve a full spectrum of values of entanglement and discord, extension of our results to a general two-qubit mixed state is still an open problem. In particular, our clean results on MMM states provide a strong motivation for validity of our observation for general mixed states that should be considered in future works. Moreover, the success of our conditional extremal analysis in studying the role of entanglement and discord suggests a similar study for other quantum information protocols.

\section*{Acknowledgments}
We would like to thank A. Ramezanpour for valuable discussions which led to the improvement of our arguments.

\appendix
\section{Calculation of the limit\label{appen}}
	We compute the limit
	\begin{equation}
		\lim\limits_{(c_1,c_2,c_3)\rightarrow 0} \frac{QA}{\delta_{in}}
		\label{eq:applim}
	\end{equation}
	where $\delta_{in}$ and $QA$ refer to initial discord and quantum advantage, respectively. Using $\delta_{enc}= p_0 \delta^{(0)}-\bar{\delta}$ for $\rho^{(0)}$ and $\bar{\rho}$ which are in the following form:
	\begin{equation}
		\rho^{(0)} =\frac{1}{4}\left( \mathds{1}\otimes\mathds{1} + \eta\sum_{i=1}^{3}c_i\sigma_i^A\otimes\sigma_i^B \right)\label{cal41},
	\end{equation}
	
	\begin{align}	
		\bar{\rho} &= \frac{1}{4}\left( \mathds{1}\otimes\mathds{1} + p_0\eta\sum_{i=1}^{3}c_i\sigma_i^A\otimes\sigma_i^B \right),\label{cal43app}
	\end{align}
	we use Eq. (\ref{anadis}) to find an expression for quantum discords corresponding to the above states.
	
	First, for the case where $\max(|c_1| , |c_2| , |c_3|) =|c_1| $, we put $\eta c_i$ instead of $c_i$. Using Eq. (\ref{anadis}), we can calculate discord for the states in Eqs. (\ref{cal41}) and (\ref{cal43app}) analytically.
	On the other hand, we have:
	\begin{equation}
		\frac{QA}{\delta_{in}}
		=\frac{\delta_{enc}}{\delta_{in}}
		=\frac{p_0\delta^{(0)} - \bar{\delta}}{\delta_{in}},
	\end{equation}
	and it is obvious that the limit (\ref{eq:lim}) is:
	\begin{equation}
		\lim\limits_{(c_1,c_2,c_3)\rightarrow 0} \frac{p_0\delta^{(0)} - \bar{\delta}}{\delta_{in}} = \frac{0}{0}.\label{limm}
	\end{equation}
	
	We need to expand the discord for the mentioned states.
	The only exceptional paths are single-axis approaches (e.g. 
			$c_2=c_3=0$),  where $\delta_{in}=0$ and the ratio is undefined; such paths are excluded, and the limit is taken over paths with $\delta_{in}\neq0$.
			
			Let $x_1=\eta c_1$, $x_2=\eta c_2$ and $x_3=\eta c_3$, and 
			\begin{equation}
				\begin{aligned}
					y_1 &= - x_1 - x_2 - x_3\\
					y_2 &= - x_1 + x_2 + x_3\\
					y_3 &= + x_1 - x_2 + x_3\\
					y_4 &= + x_1 + x_2 - x_3\\
				\end{aligned}
			\end{equation}
			So the four eigenvalues are $ \lambda_k=(1+y_k)/4$ and the four entropy terms are
			\begin{equation}
				\frac{1}{4}(1+y_k) \log_2 \frac{1}{4}(1+y_k). 
			\end{equation}
			
			We know that
			\begin{equation}
				\log_2 \frac{1}{4}(1+y_k) = -2 + \log_2 (1+y_k),
			\end{equation}
			so each eigenvalue term contributes
			\begin{align}
				&\frac{1}{4}(1+y_k) \log_2 \frac{1}{4}(1+y_k)\nonumber\\
				&= \frac{1}{4}(1+y_k) \bigg( -2+\frac{y_k}{\ln 2}\nonumber\\ &\;\;\;\;\;\;-\frac{y_k^2}{2\ln 2} + \frac{y_k^3}{3\ln 2} - \frac{y_k^4}{4\ln 2} + O(y_k^5) \bigg)\nonumber\\
				&=-\frac{1}{2}
				+\left(-\frac{1}{2}+\frac{1}{4\ln 2}\right)y_k\nonumber\\
				&\;\;\;\;\;\;+\frac{y_k^2}{8\ln 2}
				-\frac{y_k^3}{24\ln 2}
				+\frac{y_k^4}{48\ln 2}
				+O(y_k^5).
				\label{a14}
			\end{align}
			For the four values of $y_k$, we have
			\begin{equation}
				\sum_{k=1}^{4}y_k=0,
			\end{equation}
			so the linear terms cancel. Furthermore,
			\begin{equation}
				\sum_{k=1}^{4}y_k^2
				=
				4(x_1^2+x_2^2+x_3^2),
				\label{a16}
			\end{equation}
			and the cubic terms are
			\begin{align}
				\sum_{k=1}^{4}y_k^3
				&=
				(-x_1-x_2-x_3)^3
				+(-x_1+x_2+x_3)^3
				\nonumber\\
				&\quad
				+(x_1-x_2+x_3)^3
				+(x_1+x_2-x_3)^3
				\nonumber\\
				&=-24x_1x_2x_3.
				\label{a18}
			\end{align}
			The fourth-order terms satisfy
			\begin{equation}
				\begin{aligned}
				\sum_{k=1}^{4}y_k^4
				=4&(x_1^4+x_2^4+x_3^4)\\
				&+24(x_1^2x_2^2+x_1^2x_3^2+x_2^2x_3^2).
				\end{aligned}
				\label{a19}
			\end{equation}
			Consequently, the four eigenvalue terms yield
			\begin{equation}
			\begin{aligned}
				\sum_{k=1}^{4}\lambda_k\log_2\lambda_k
				={}&-2
				+\frac{x_1^2+x_2^2+x_3^2}{2\ln 2}\\
				&+\frac{x_1x_2x_3}{\ln 2}
				+\frac{x_1^4+x_2^4+x_3^4}{12\ln 2}\\
				&+\frac{x_1^2x_2^2+x_1^2x_3^2+x_2^2x_3^2}{2\ln 2}
				+O(\|x\|^5).
				\label{a17}
			\end{aligned}
			\end{equation}
			For the classical-correlation term, regarding the Eq. (\ref{classical}) where $\zeta = |x_1|$, we have
			\begin{equation}
				-\frac{1-|x_1|}{2}\log_2(1-|x_1|)
				-\frac{1+|x_1|}{2}\log_2(1+|x_1|).
				\label{cal66}
			\end{equation}
			We expand the complete expression. This gives
			\begin{equation}
				-\frac{x_1^2}{2\ln 2}
				-\frac{x_1^4}{12\ln 2}
				+O(x_1^6).
				\label{a22}
			\end{equation}
			Therefore, combining Eqs. (\ref{a17}) and (\ref{a22}), the discord of
			$\rho^{(0)}$ is
			
			\begin{equation}
			\begin{aligned}
				\delta^{(0)}
				={}&
				\frac{\eta^2}{2\ln 2}(c_2^2+c_3^2)
				+\frac{\eta^3}{\ln 2}c_1c_2c_3\\
				&+\frac{\eta^4}{12\ln 2}(c_2^4+c_3^4)
				+\frac{\eta^4}{2\ln 2}
				(c_1^2c_2^2+c_1^2c_3^2+c_2^2c_3^2)\\
				&+O(\|c\|^5).
				\label{a23}
			\end{aligned}
			\end{equation}
						
			Similarly, for $\bar{\rho}$,
			\begin{equation}
				\begin{aligned}
				\bar{\delta}
				={}&
				\frac{p_0^2\eta^2}{2\ln 2}(c_2^2+c_3^2)
				+\frac{p_0^3\eta^3}{\ln 2}c_1c_2c_3\\
				&+\frac{p_0^4\eta^4}{12\ln 2}(c_2^4+c_3^4)
				+\frac{p_0^4\eta^4}{2\ln 2}
				(c_1^2c_2^2+c_1^2c_3^2+c_2^2c_3^2)\\
				&+O(\|c\|^5).
				\label{a24}
				\end{aligned}
			\end{equation}
			
			For the initial state, corresponding to $\eta=1$,
			\begin{equation}
				\begin{aligned}
				\delta_{in}
				={}&
				\frac{1}{2\ln 2}(c_2^2+c_3^2)
				+\frac{1}{\ln 2}c_1c_2c_3\\
				&+\frac{1}{12\ln 2}(c_2^4+c_3^4)
				+\frac{1}{2\ln 2}
				(c_1^2c_2^2+c_1^2c_3^2+c_2^2c_3^2)\\
				&+O(\|c\|^5).
				\label{a25}
				\end{aligned}
			\end{equation}
			
			To evaluate the limit, we first collect the leading contributions to
			$p_0\delta^{(0)}-\bar{\delta}$. From Eqs. (\ref{a23}) and
			(\ref{a24}), we obtain
			\begin{equation}
				\begin{aligned}
				p_0\delta^{(0)}-\bar{\delta}
				={}&
				\frac{p_0\eta^2(1-p_0)}{2\ln 2}(c_2^2+c_3^2)\\
				&+\frac{p_0\eta^3(1-p_0^2)}{\ln 2}c_1c_2c_3\\
				&+\frac{p_0\eta^4(1-p_0^3)}{12\ln 2}
				(c_2^4+c_3^4)\\
				&+\frac{p_0\eta^4(1-p_0^3)}{2\ln 2}
				(c_1^2c_2^2+c_1^2c_3^2+c_2^2c_3^2)\\
				&+O(\|c\|^5).
				\label{a26}
				\end{aligned}
			\end{equation}
			
			The remainder terms in Eqs.~(\ref{a23})--(\ref{a25}) are also $o(c_2^2+c_3^2)$.
			Indeed, in the region $|c_1|=\max_i|c_i|$, $\delta$ coincides with a function analytic near $c=0$ and is invariant under flipping the signs of any two of the $c_i$ (this merely permutes the $y_k$). Hence every term $c_1^a c_2^b c_3^c$ in its Taylor series has $a,b,c$ of equal parity. Moreover $\delta=0$ when $c_2=c_3=0$, so (whether $b,c$ are both even or both odd) $b+c\ge 2$ .
			
			Each term is then bounded by $\|c\|^{a+b+c-2}(c_2^2+c_3^2)$: writing
			$m=\max(|c_2|,|c_3|)$, so that $m^2\le c_2^2+c_3^2$ and $m\le\|c\|$, and
			using $|c_1|\le\|c\|$, we have
			\begin{equation}
			\begin{aligned}
				|c_1^ac_2^bc_3^c|
				\le& \|c\|^{a}\,m^{b+c}\\
				&= \|c\|^{a}\,m^{b+c-2}\,m^{2}\\
				&\;\;\;\;\le \|c\|^{a}\,\|c\|^{b+c-2}\,(c_2^2+c_3^2)\\
				&\;\;\;\;\;\;\;= \|c\|^{a+b+c-2}(c_2^2+c_3^2),
			\end{aligned}
			\end{equation}
			where the last power $m^{b+c-2}\le\|c\|^{b+c-2}$ is well defined since
			$b+c\ge2$. So all terms beyond quadratic order are
			$O\big(\|c\|(c_2^2+c_3^2)\big)=o(c_2^2+c_3^2)$.
			
			Therefore, for approaches with $\delta_{in}\neq0$, the quadratic term
			dominates and
			\begin{equation}
				p_0\delta^{(0)}-\bar{\delta}
				=
				\frac{p_0\eta^2(1-p_0)}{2\ln 2}
				(c_2^2+c_3^2)
				+o(c_2^2+c_3^2).
				\label{a30}
			\end{equation}
			where $o(c_2^2+c_3^2)$ denotes that the remaining terms vanish faster than $c_2^2+c_3^2$ itself. From Eq. (\ref{a25}), we likewise have
			\begin{equation}
				\delta_{in}
				=
				\frac{1}{2\ln 2}(c_2^2+c_3^2)
				+o(c_2^2+c_3^2).
				\label{a31}
			\end{equation}
			Consequently,
			\begin{equation}
				\lim\limits_{(c_1,c_2,c_3)\rightarrow0}
				\frac{p_0\delta^{(0)}-\bar{\delta}}{\delta_{in}}
				=
				p_0\eta^2(1-p_0).
				\label{a32}
			\end{equation}
			The same result follows for the cases
			$|c_2|=\max_i|c_i|$ and $|c_3|=\max_i|c_i|$ by permutation of the
			indices.


\begin{thebibliography}{99}
	
	\bibitem{chitambar2019quantum}
	E.~Chitambar and G.~Gour,
	``Quantum resource theories,''
	\emph{Rev. Mod. Phys.} \textbf{91}, 025001 (2019).
	
	\bibitem{fanchini2017lectures}
	F.~F.~Fanchini, D.~O.~S.~Pinto, and G.~Adesso,
	\emph{Lectures on general quantum correlations and their applications}
	(Springer, 2017).
	
	\bibitem{adesso2016measures}
	G.~Adesso, T.~R.~Bromley, and M.~Cianciaruso,
	``Measures and applications of quantum correlations,''
	\emph{J. Phys. A: Math. Theor.} \textbf{49}, 473001 (2016).
	
	\bibitem{anshu2018quantifying}
	A.~Anshu, M.-H.~Hsieh, and R.~Jain,
	``Quantifying resources in general resource theory with catalysts,''
	\emph{Phys. Rev. Lett.} \textbf{121}, 190504 (2018).
	
	\bibitem{vedral2003classical}
	V.~Vedral,
	``Classical correlations and entanglement in quantum measurements,''
	\emph{Phys. Rev. Lett.} \textbf{90}, 050401 (2003).
	
	\bibitem{vedral1997quantifying}
	V.~Vedral, M.~B.~Plenio, M.~A.~Rippin, and P.~L.~Knight,
	``Quantifying entanglement,''
	\emph{Phys. Rev. Lett.} \textbf{78}, 2275 (1997).
	
	\bibitem{horodecki2001entanglement}
	M.~Horodecki,
	``Entanglement measures,''
	\emph{Quantum Inf. Comput.} \textbf{1}, 3--26 (2001).
	
	\bibitem{lee2025quantum}
	S.~Lee, C.~Noh, and J.~Park,
	``Quantum discord is not extremalized by Gaussian states: S. Lee et al.,''
	\emph{Quantum Inf. Process.} \textbf{24}, 80 (2025).
	
	\bibitem{liu2015general}
	F.~Liu, G.-J.~Tian, Q.-Y.~Wen, and F.~Gao,
	``General bounds for quantum discord and discord distance,''
	\emph{Quantum Inf. Process.} \textbf{14}, 1333--1341 (2015).
	
	\bibitem{rau2018calculation}
	A.~R.~P.~Rau,
	``Calculation of quantum discord in higher dimensions for X-and other specialized states: ARP Rau,''
	\emph{Quantum Inf. Process.} \textbf{17}, 216 (2018).
	
	\bibitem{yurischev2015quantum}
	M.~A.~Yurischev,
	``On the quantum discord of general X states,''
	\emph{Quantum Inf. Process.} \textbf{14}, 3399--3421 (2015).
	
	\bibitem{henderson2001classical}
	L.~Henderson and V.~Vedral,
	``Classical, quantum and total correlations,''
	\emph{J. Phys. A: Math. Gen.} \textbf{34}, 6899 (2001).
	
	\bibitem{ollivier2001quantum}
	H.~Ollivier and W.~H.~Zurek,
	``Quantum discord: a measure of the quantumness of correlations,''
	\emph{Phys. Rev. Lett.} \textbf{88}, 017901 (2001).
	
	\bibitem{modi2012classical}
	K.~Modi, A.~Brodutch, H.~Cable, T.~Paterek, and V.~Vedral,
	``The classical-quantum boundary for correlations: quantum discord and related measures,''
	\emph{Rev. Mod. Phys.} \textbf{84}, 1655--1707 (2012).
	
	\bibitem{qasimi2011comparison}
	A.~Al--Qasimi and D.~F.~V.~James,
	``Comparison of the attempts of quantum discord and quantum entanglement to capture quantum correlations,''
	\emph{Phys. Rev. A} \textbf{83}, 032101 (2011).
	
	\bibitem{degen2017quantum}
	C.~L.~Degen, F.~Reinhard, and P.~Cappellaro,
	``Quantum sensing,''
	\emph{Rev. Mod. Phys.} \textbf{89}, 035002 (2017).
	
	\bibitem{ekert1998quantum}
	A.~Ekert and R.~Jozsa,
	``Quantum algorithms: entanglement--enhanced information processing,''
	\emph{Phil. Trans. R. Soc. A} \textbf{356}, 1769--1782 (1998).
	
	\bibitem{steane1998quantum}
	A.~Steane,
	``Quantum computing,''
	\emph{Rep. Prog. Phys.} \textbf{61}, 117 (1998).
	
	\bibitem{jozsa2003role}
	R.~Jozsa and N.~Linden,
	``On the role of entanglement in quantum-computational speed-up,''
	\emph{Proc. R. Soc. A} \textbf{459}, 2011--2032 (2003).
	
	\bibitem{pan2001entanglement}
	J.-W.~Pan, C.~Simon, Č.~Brukner, and A.~Zeilinger,
	``Entanglement purification for quantum communication,''
	\emph{Nature} \textbf{410}, 1067--1070 (2001).
	
	\bibitem{pirandola2015advances}
	S.~Pirandola, J.~Eisert, C.~Weedbrook, A.~Furusawa, and S.~L.~Braunstein,
	``Advances in quantum teleportation,''
	\emph{Nat. Photonics} \textbf{9}, 641--652 (2015).
	
	\bibitem{sherson2006quantum}
	J.~F.~Sherson, H.~Krauter, R.~K.~Olsson, B.~Julsgaard, K.~Hammerer, I.~Cirac, and E.~S.~Polzik,
	``Quantum teleportation between light and matter,''
	\emph{Nature} \textbf{443}, 557--560 (2006).
	
	\bibitem{vaidman1994teleportation}
	L.~Vaidman,
	``Teleportation of quantum states,''
	\emph{Phys. Rev. A} \textbf{49}, 1473 (1994).
	
	\bibitem{bennett1993teleporting}
	C.~H.~Bennett, G.~Brassard, C.~Crépeau, R.~Jozsa, A.~Peres, and W.~K.~Wootters,
	``Teleporting an unknown quantum state via dual classical and Einstein-Podolsky-Rosen channels,''
	\emph{Phys. Rev. Lett.} \textbf{70}, 1895 (1993).
	
	\bibitem{jennewein2000quantum}
	T.~Jennewein, C.~Simon, G.~Weihs, H.~Weinfurter, and A.~Zeilinger,
	``Quantum cryptography with entangled photons,''
	\emph{Phys. Rev. Lett.} \textbf{84}, 4729 (2000).
	
	\bibitem{bennett1992quantum}
	C.~H.~Bennett, G.~Brassard, and A.~K.~Ekert,
	``Quantum cryptography,''
	\emph{Sci. Am.} \textbf{267}, 50--57 (1992).
	
	\bibitem{tittel2000quantum}
	W.~Tittel, J.~Brendel, H.~Zbinden, and N.~Gisin,
	``Quantum cryptography using entangled photons in energy-time Bell states,''
	\emph{Phys. Rev. Lett.} \textbf{84}, 4737 (2000).
	
	\bibitem{amico2008entanglement}
	L.~Amico, R.~Fazio, A.~Osterloh, and V.~Vedral,
	``Entanglement in many-body systems,''
	\emph{Rev. Mod. Phys.} \textbf{80}, 517--576 (2008).
	
	\bibitem{nielsen2010quantum}
	M.~A.~Nielsen and I.~L.~Chuang,
	\emph{Quantum computation and quantum information}
	(Cambridge University Press, 2010).
	
	\bibitem{datta2008quantum}
	A.~Datta, A.~Shaji, and C.~M.~Caves,
	``Quantum discord and the power of one qubit,''
	\emph{Phys. Rev. Lett.} \textbf{100}, 050502 (2008).
	
	\bibitem{madhok2013quantum}
	V.~Madhok and A.~Datta,
	``Quantum discord as a resource in quantum communication,''
	\emph{Int. J. Mod. Phys. B} \textbf{27}, 1345041 (2013).
	
	\bibitem{pirandola2014quantum}
	S.~Pirandola,
	``Quantum discord as a resource for quantum cryptography,''
	\emph{Sci. Rep.} \textbf{4}, 6956 (2014).
	
	\bibitem{brodutch2013discord}
	A.~Brodutch,
	``Discord and quantum computational resources,''
	\emph{Phys. Rev. A} \textbf{88}, 022307 (2013).
	
	\bibitem{datta2011quantum}
	A.~Datta and A.~Shaji,
	``Quantum discord and quantum computing—an appraisal,''
	\emph{Int. J. Quantum Inf.} \textbf{9}, 1787--1805 (2011).
	
	\bibitem{gu2012observing}
	M.~Gu, H.~M.~Chrzanowski, S.~M.~Assad, T.~Symul, K.~Modi, T.~C.~Ralph, V.~Vedral, and P.~K.~Lam,
	``Observing the operational significance of discord consumption,''
	\emph{Nat. Phys.} \textbf{8}, 671--675 (2012).
	
	\bibitem{lloyd2008enhanced}
	S.~Lloyd,
	``Enhanced sensitivity of photodetection via quantum illumination,''
	\emph{Science} \textbf{321}, 1463--1465 (2008).
	
	\bibitem{weedbrook2016discord}
	C.~Weedbrook, S.~Pirandola, J.~Thompson, V.~Vedral, and M.~Gu,
	``How discord underlies the noise resilience of quantum illumination,''
	\emph{New J. Phys.} \textbf{18}, 043027 (2016).
	
	\bibitem{kim2023entanglement}
	M.~Kim, M.-R.~Hwang, E.~Jung, and D.~Park,
	``Is entanglement a unique resource in quantum illumination? M. Kim et al.,''
	\emph{Quantum Inf. Process.} \textbf{22}, 98 (2023).
	
	\bibitem{yung2020one}
	M.-H.~Yung, F.~Meng, X.-M.~Zhang, and M.-J.~Zhao,
	``One-shot detection limits of quantum illumination with discrete signals,''
	\emph{npj Quantum Inf.} \textbf{6}, 75 (2020).
	
	\bibitem{jo2021quantum}
	Y.~Jo, T.~Jeong, J.~Kim, D.~Y.~Kim, Y.~S.~Ihn, Z.~Kim, and S.-Y.~Lee,
	``Quantum illumination with asymmetrically squeezed two-mode light,''
	\emph{arXiv preprint arXiv:2103.17006} (2021).
	
	\bibitem{bradshaw2017overarching}
	M.~Bradshaw, S.~M.~Assad, J.~Y.~Haw, S.-H.~Tan, P.~K.~Lam, and M.~Gu,
	``Overarching framework between Gaussian quantum discord and Gaussian quantum illumination,''
	\emph{Phys. Rev. A} \textbf{95}, 022333 (2017).
	
	\bibitem{maziero2009classical}
	J.~Maziero, L.~C.~Celeri, R.~M.~Serra, and V.~Vedral,
	``Classical and quantum correlations under decoherence,''
	\emph{Phys. Rev. A} \textbf{80}, 044102 (2009).
	
	\bibitem{pozzobom2019preparing}
	M.~B.~Pozzobom and J.~Maziero,
	``Preparing tunable Bell-diagonal states on a quantum computer,''
	\emph{Quantum Inf. Process.} \textbf{18}, 142 (2019).
	
	\bibitem{hou2018geometric}
	J.-X.~Hou, Y.-L.~Su, S.-Y.~Liu, X.-H.~Wang, and W.-L.~Yang,
	``Geometric structure of quantum resources for Bell-diagonal states: J.-X. Hou et al.,''
	\emph{Quantum Inf. Process.} \textbf{17}, 184 (2018).
	
	\bibitem{lang2010quantum}
	M.~D.~Lang and C.~M.~Caves,
	``Quantum discord and the geometry of Bell-diagonal states,''
	\emph{Phys. Rev. Lett.} \textbf{105}, 150501 (2010).
	
	\bibitem{virmani2000ordering}
	S.~Virmani and M.~B.~Plenio,
	``Ordering states with entanglement measures,''
	\emph{Phys. Lett. A} \textbf{268}, 31--34 (2000).
	
	\bibitem{miranowicz2004comparative}
	A.~Miranowicz and A.~Grudka,
	``A comparative study of relative entropy of entanglement, concurrence and negativity,''
	\emph{J. Opt. B} \textbf{6}, 542--548 (2004).
	
	\bibitem{bromley2014unifying}
	T.~R.~Bromley, M.~Cianciaruso, R.~L.~Franco, and G.~Adesso,
	``Unifying approach to the quantification of bipartite correlations by Bures distance,''
	\emph{J. Phys. A: Math. Theor.} \textbf{47}, 405302 (2014).
	
	\bibitem{dakic2010necessary}
	B.~Dakić, V.~Vedral, and Č.~Brukner,
	``Necessary and sufficient condition for nonzero quantum discord,''
	\emph{Phys. Rev. Lett.} \textbf{105}, 190502 (2010).
	
	\bibitem{quan2016steering}
	Q.~Quan, H.~Zhu, S.-Y.~Liu, S.-M.~Fei, H.~Fan, and W.-L.~Yang,
	``Steering Bell-diagonal states,''
	\emph{Sci. Rep.} \textbf{6}, 22025 (2016).
	
	\bibitem{wootters1998entanglement}
	W.~K.~Wootters,
	``Entanglement of formation of an arbitrary state of two qubits,''
	\emph{Phys. Rev. Lett.} \textbf{80}, 2245 (1998).
	
	\bibitem{xu2015quantum}
	P.~Xu, T.~Wu, and L.~Ye,
	``The quantum correlations of the werner state under quantum decoherence,''
	\emph{Int. J. Theor. Phys.} \textbf{54}, 1958--1967 (2015).
	
	\bibitem{holevo2002capacity}
	A.~S.~Holevo,
	``The capacity of the quantum channel with general signal states,''
	\emph{IEEE Trans. Inf. Theory} \textbf{44}, 269--273 (2002).
	
	\bibitem{fuchs1994ensemble}
	C.~A.~Fuchs and C.~M.~Caves,
	``Ensemble-dependent bounds for accessible information in quantum mechanics,''
	\emph{Phys. Rev. Lett.} \textbf{73}, 3047 (1994).
	
	\bibitem{fuchs1996distinguishability}
	C.~A.~Fuchs,
	``Distinguishability and accessible information in quantum theory,''
	\emph{arXiv preprint quant-ph/9601020} (1996).
	
	\bibitem{ray2019maximum}
	S.~Ray, J.~Schneeloch, C.~C.~Tison, and P.~M.~Alsing,
	``Maximum advantage of quantum illumination,''
	\emph{Phys. Rev. A} \textbf{100}, 012327 (2019).
	
	\bibitem{piani2012problem}
	M.~Piani,
	``Problem with geometric discord,''
	\emph{Phys. Rev. A} \textbf{86}, 034101 (2012).
	
\end{thebibliography}
\end{document}